\documentstyle[prl,aps,epsfig]{revtex}

\newcommand{\beq}{\begin{equation}}
\newcommand{\eeq}{\end{equation}}
\newcommand{\beqar}{\begin{eqnarray}}
\newcommand{\eeqar}{\end{eqnarray}}
\newcommand{\bfig}{\begin{figure}}
\newcommand{\efig}{\end{figure}}
\newcommand{\bd}{\begin{itemize}} %description}}
\newcommand{\ed}{\end{itemize}} %description}}
\newcommand{\bc}{\begin{center}}
\newcommand{\ec}{\end{center}}
\newcommand{\be}{\begin{equation}}
\newcommand{\ee}{\end{equation}}

\newcommand{\ba}{\begin{array}}
\newcommand{\ea}{\end{array}}

\newcommand{\set}[2]{\newcommand{#1}{#2}}
\set{\pa}{\partial \over \partial\, }
\set{\leftvector}{\stackrel{\leftarrow}{\partial }}
\set{\rightvector}{\stackrel{\rightarrow}{\partial }}
\begin{document}
%************************************************************************
\twocolumn[\hsize\textwidth\columnwidth\hsize
           \csname @twocolumnfalse\endcsname
%-----------------------------------------------------------------------

\title{The size of two-body weakly bound objects :
short versus long range potentials}
\author{R. Lombard$^{a}$ and C. Volpe$^{a,b}$}
%\date{}
\address{$^{a)}$
  Groupe de Physique Th\'{e}orique, Institut de Physique Nucl\'{e}aire,
F-91406 Orsay Cedex, France \\
$^{b)}$
Institut f\"ur Theoretische Physik der Universit\"at, Philosophenweg
19, D-69120 Heidelberg, Germany \\}

\maketitle

\begin{abstract}
The variation of the size of two-body objects is investigated, as
the separation energy approaches zero, 
with both long range potentials and short range  potentials
having a repulsive core.  
It is shown that long range potentials can also give rise
to very extended systems.
The asymptotic laws 
derived for states with angular momentum $\ell=1,2$
differ 
from the ones obtained with short range potentials.
The sensitivity of the asymptotic laws on the shape and
length of short range potentials defined by two and three
parameters is studied. 
These ideas as well as the transition from the short 
to the long range regime for the $\ell=0$ case  are illustrated
using the Kratzer potential.

\end{abstract}

\vskip2pc
]

\noindent
The study of the properties 
 of weakly bound systems
has found a renewed
interest after the discovery of halo nuclei in nuclear physics 
\cite{Rii,HJJ,TAN}
as
well as loosely bound dimers in molecular physics \cite{STO,KLM}.
These systems have very large mean square radii and
small separation energies. 
They can be treated as two (or three)-body systems interacting
through a potential. In fact,
the separation energy of 
the one or two nucleons forming the halo is so small that their
degrees of freedom can be separated from those of the nucleons
constituents of the core. In the case of diffuse diatomic molecules,
the situation is even better, the separation energy being several orders of
magnitude less than the ionization potential of each constitutent.

In this Letter, we want to discuss in very general terms 
the size of two-body weakly bound objects. 
Short and long range repulsive forces have been studied
up to now in the context of nuclear and molecular physics.
In particular, the variation of the size of these quantum systems
as the energy approches zero (asymptotic 
laws) 
has been actively investigated \cite{FJR,RFJ}.

There is one more class of potentials which can give
rise to very extended systems in the zero energy limit,
namely long-range potentials 
for which this limit is attained when the
deepness of the potential goes to zero.
It is the purpose of this paper to discuss these long-range
potentials and show that they predict a variation of the
size of the system, in the zero energy limit, different
from short-range potentials.
We will compare the two behaviours and discuss the transition
between the short and long range regime.

Concerning short range potentials, 
we will discuss short range
potentials presenting a repulsive core.
The presence of a repulsive core 
in the interacting potential should actually be included to 
modelize the Pauli exclusion principle.
We will show how the presence of the core modifies the 
asymptotic 
laws found for
short range potentials, 
%cri
discussed up to now, which are
%end
defined by only two parameters, namely the deepness
of the potential and the range.

The Letter is structured as follows.
We will first show that asymptotic 
laws for states of any
angular momentum $l$ can be derived
starting from the Schr\"odinger equation, without making any specific
reference to the potential.
Then we will discuss how we can derive these 
laws by using
the asymptotic behaviour of the wave functions both for short and
long range potentials.
Finally we will use the Kratzer potential to illustrate the energy
dependence of the mean square radius for long range potentials, 
the influence of the core in the case of short range potentials
(obtained from the Kratzer potential by cutting it at some distance)
and the transition from the short to the long range regime.
We will use the results obtained for a square well for a comparison
with short range potentials defined by 
%cri
only
%end
two parameters.   

In this work, we will consider only spherical symmetric potentials. The spin
degrees of freedom are ignored.  
In this case, the
radial Schr\"odinger equation ($\hbar = 2m = 1$) is given by
\begin{equation}\label{1}
[ - \frac{\partial ^2}{\partial r^2} - \frac{2}{r} \frac{\partial}{\partial r}
+ \frac{\ell(\ell+1)}{r^2} + \lambda w(r) ] 
\Psi_{\ell}(r)= E_{\ell}\Psi_{\ell}(r) . 
\end{equation}
%cri
where $\lambda$ scales the deepness of the potential
$w(r)$.
%end
Since we consider only the lowest states of each angular momentum (the
wavefunction has no node), they are simply labelled by $\ell$.

The first very general result obtained from eq.(\ref{1}), 
without any specific reference to the potential, is a kind of
Heisenberg relation \cite{GAM} :
\be\label{2}
<r^2>_{\ell} \ \ \geq \ \ \frac{(2 \ell + 3)^2}{4 <T>_{\ell}} \ ,
\ee
\noindent
relating the
 rms radius to the average kinetic energy.
%%%%%%%%
For confining potentials, we have $E_{\ell} \geq <T>_{\ell}$, the equality
being reached as $\lambda \rightarrow 0$. Moreover, for power-law potentials
($w(r) = r^{\alpha}, \ \alpha > -2$) (or superpositions of them) the virial
theorem
\be
<T>_{\ell} = \frac{\lambda}{2} \int \rho_{\ell}(r) r [\frac{\partial w}{\partial
r}] d^3r \ . \ee
ensures that $<T>_{\ell} \propto <V>_{\ell}$ and thus
$<T>_{\ell} \propto E_{\ell}$.
Consequently, for this large class of potentials,
the inequality (2) readily tells us that the rms radius is diverging with
$1/E_{\ell}$ for all $\ell$ as $E_{\ell} \rightarrow 0$, which means 
$\lambda \rightarrow 0$. 

%We conjecture this conclusion to be valid for any
%long range potential.

%%%%%%%%%%%%%%%%%%%%%

No such a general prediction can be made in the case of short range
potentials, because in this case the zero energy limit is obtained when
the coupling $\lambda$ tends to a critical finite value $\lambda_c$ 
\cite{LAL}. 
 
In the case $\ell=0$, another prediction is given by  
the Bertlmann-Martin inequality \cite{BAM} which concerns the ground
state radius and yields
\be\label{4}
<r^2>_0 \leq  \frac{3}{E_1 - E_0}    \ . 
\ee
where $E_0$ and $E_1$ are the energies of the ground and first
excited states, respectively. 
Noting
that $E_1 - E_0 >0$, we can write 
$E_1 = \varphi E_0\  $ and get
\be\label{4bis}
<r^2>_0 \leq \frac{3}{(\varphi-1) E_0} \ \ , \ \  \ \varphi -1 >0 \ . \ee
Consequently, for $\ell =0$ the asymptotic
behaviour of the rms radius as the energy tends to zero is given by this
expression for all, short or long range, potentials.

In order to specify the value of $\varphi$ and obtain 
predictions for $\ell \not= 0$, it is necessary to introduce
assumptions concerning the wavefunctions. 
%Though quite general, they may be
%valid only in a restricted domain of energies. 

Let's first discuss short range potentials, characterized by a finite
range $R_0$.
Since the potential becomes negligible beyond $R_0$,
the asymptotic form of the wavefunctions is related to the 
spherical
Hankel functions
\be \label{6}
\Psi_{\ell}(r) \ \approx \ \frac{e^{-\mu r}}{r^{\ell}} \ .\ee
As $\mu \rightarrow 0$ the energy tends to zero. Therefore, 
as long as the 
contribution from the inner part of the 
wavefunction 
is negligible, the rms radius is
given by
\be\label{7}
<r^2>_{\ell} \approx \int_{R_0}^{\infty} e^{-2\mu r} r^{2 - 2\ell}dr/
\int_{R_0}^{\infty} e^{-2\mu r} r^{- 2\ell}dr \ . \ee
With this approximation, we get the relations derived by Riisager, Jensen
and M{\"o}ller \cite{RJM} :
%cri
\be\label{8}
<r^2>_0 \approx \frac{c_0}{|E_0|} \ ; 
\ <r^2>_1 \approx \frac{c_1 \ R_0}{\sqrt{|E_1|}} \ ; \
<r^2>_2 \approx {c_2 \ R_0^2}  \ ,\ee
%end

Intuitively, the above assumptions imply
the constants $c_{\ell}$ to be independent of the potential. However this
statement is not exact and in general $c_{\ell} = c_{\ell}(w) $ is expected.
For $\ell \geq 3$ the rms radius behaves in the same way as for $\ell = 2$, 
namely it
tends to a constant as the energy approaches zero. 

%Although the derivation
%is really valid asymptotically, this approximation is reasonable on a 
%wider energy range, which has to be specified.

Coming back to the inequality (\ref{4bis}),
in the case the $\ell = 1$ state is in the
continuum, $E_1$ is set to zero  (or $\varphi = 0$). (Note that this
procedure is not valid for confining potentials, for which
$E_{\ell} > 0$.)  
Using the asymptotic 
properties of the s-state
wavefunction leads to an absolute lower bound \cite{LKL}
\be \label{5}
<r^2>_0 \geq \frac{1}{2|E_0|}   \ . 
\ee
The equality is reached as $-E_0 \rightarrow 0$, independently on the
potential. As a consequence, in eq.(\ref{8}), $c_0 = 1/2$ is a firm
prediction, independently of the shape of the short range potential.

To get a practical insight on how much $c_{\ell}$ depends on $w$,
the asymptotic laws (\ref{8}) have also been studied numerically 
for potentials depending  only on two parameters, the deepness and a
typical length $R_0$. In particular, 
the square well
and the gaussian potentials have been used  \cite{RJM}. 
Their typical lengths
are the
square well radius and the range of the gaussian, respectively. 
In \cite{RJM}, it is shown that the coefficients $c_{\ell}$
seem to be not very sensitive to the particular shape of the potential.

In general, $<r^2>_{\ell}=f(E,R_0,w)$, that is the mean square radii depend
on the energy of the state, the typical length $R_0$ and the shape
of the potential. Eq.(\ref{8}) explicitly give the dependence of
$<r^2>_{\ell}$ on $R_0$ for short range potentials defined by two parameters.
This can be seen directly from the  Schr\"odinger equation,
by means of the
change of variable $x =r/R_0$. As a result,
$<x^2>_{\ell}=f'(\epsilon,w)$ where $\epsilon=E_{\ell}R_0^2$. 
As soon as the potential contains more than two parameters, the scaling
cannot be exact, although it may constitute a good approximation under some
circonstances. As a consequence of this {\it quasi}-scaling, 
the $<r^2>_{\ell}$ may 
depend on the length
of the potential in a way similar to (\ref{8}). 
 Such a situation may occur in particular when the potential
has a short range repulsive component.

Let us now discuss the asymptotic laws 
for long range potentials, using the behaviour of the wave functions. 
Contrary to short range potentials, 
the wavefunction is in this case
confined inside
the potential. Use can be made of the spherical Bessel function of the
first kind $j_{\ell}(kr)$, with $k \rightarrow 0$, cutting the integrals 
at the first zero. This procedure yields right away
\be\label{9}
<r^2>_{\ell} \approx  \frac{c_{\ell}(w)}{|E_{\ell}|} \ .\ee
The derivation is very crude, and merely confirms the above statement of
eqs.(\ref{2}) and (\ref{4})-(\ref{4bis}). 

Relations (\ref{2}),
%cri
(\ref{8}) 
%end
and (\ref{9}) show that both short and long range
potentials may predict very diffuse systems in the zero energy limit.
The asymptotic 
laws obtained present similarities as well as differences.
In fact, in the case of long range potentials, the rms diverges as
$1/E_{\ell}$ not only for $\ell=0$, 
as it has been found for short range potentials,
but for 
any $\ell$. Besides,
the constants $c_{\ell}$
are sensitive to the potential $w$. 

A question one may ask about systems with halos is how to define
them. In the case of short range potentials, a natural definition
arises as it has been extensively discussed in the literature \cite{RFJ}.
A reference scale 
is needed to compare it  to the mean square radius of the system
in order to quantify the size of the halo.
This reference scale can be taken equal to the range of the short
range potential which in some cases can be related to a physical 
scale of the system,
for example the size of the core in a halo nucleus and is identified
with the outer classical turning point. An ideal halo is then defined
as a system for which the single halo particle has a very large
probability  to  be inside the classically  forbidden region.
Its properties are therefore determined by the tail of the
wavefunction
and are almost independent of the potential.

For the s-state, the above discussion refers to $E_0 \rightarrow 0$.
However, it is interesting to note that for the shell-delta potential,
$w(r)=\delta(r-R_0)$ which possesses a single bound s-level,
the particle is always in the classically forbidden region for
any finite $E_0$.

In the case of long range potentials, the definition of a halo seems
less straightforward. If the potential presents a repulsif core,
a natural reference scale is given by the size of this core which
can be related again to some physical size as, for example, that
of the core in a halo nucleus or the sum to the two atomic radii, in the case
of a diffuse molecule. So, the mean square radius of the system can 
be compared to the range of the potential core in order to quantify
the extension of the halo. Contrary to the case of short range
potentials, the wavefunction of the single halo particle will always
be confined in the classically allowed region and the probability of
finding the particle in the potential will always be one.
Finally, in the case of short range potentials, 
an ideal halo is only possible if the halo particle is
in an s-state.
In fact, for p- a d-states the centrifugal barrier plays an important
role, confining the wavefunction. This is not the case for long range
potentials, where the barrier plays no role and therefore
wavefunctions of any angular momentum can be extended.

Let us now discuss the asymptotic
laws for long and short range
potentials as well as the transition between the two in a particular
case, namely the Kratzer potential \cite{FLU} :
\be\label{10}
w(r)=-2\lambda({a \over r}-{a^2 \over {2r^2}})
\ee
where $a$ is just a scaling parameter which gives the distance at
which the potential changes from attractive to repulsive.       
This potential has a long range and a repulsive core.  
The short range case is obtained by putting
the potential to zero at a variable distance $R_{cut}$ : 
%cri
\be\label{10bis}
w_{cut}(r)=-2\lambda({a \over r}-{a^2 \over {2r^2}})\Theta(R_{cut}-r)
\ee
This cut Kratzer potential
%end
has
the peculiarity of being defined by three parameters, namely the size of
the core, the deepness of the potential $\lambda$ and the range $R_{cut}$. 

By solving the Schr\"odinger equation (\ref{1}) with (\ref{10}), 
we obtain the mean square radius for any $\ell$ :
\be\label{11}
\langle r^2 \rangle _{\ell} 
= {1 \over {2 \vert E_l \vert }} \left[ 3 + 5 
\sqrt{\gamma^2 + ({\ell}+{1 \over 2})^2} + 2 (\gamma^2 + 
({\ell}+{1 \over 2})^2)  \right]
\ee
where $\gamma^2 = (a^2 \lambda)$.
This gives, when $E_l \rightarrow 0$,   
\be\label{12}
\langle r^2 \rangle_0 \approx {3 \over {E_0}}~~~~
\langle r^2 \rangle_1 \approx {7.5 \over {E_1}}~~~~
\langle r^2 \rangle_2 \approx {14 \over {E_2}} 
\ee
As expected, $\langle r^2 \rangle _{\ell}$ have the same dependence on the
energy as predicted by (\ref{2}) and (\ref{9}). For the sake of comparison,
note that the values of
$c_{\ell}$ are the same in the case of a pure Coulomb force. For the
harmonic oscillator, $c_0$ and $c_1$ take the same values whereas $c_2 = 10.5$.

%cri
We have solved (\ref{1})  
with the cut Kratzer potential
(\ref{10bis}) 
for different values of $R_{cut}$ in order
to see explicitly how the asymptotic laws  vary
when we use a three instead of a two parameter potential. 
Particular attention has been devoted to the asymptotic region, working down
to energies of the order of $10^{-10}$. Note that such an achievement is only
possible for potentials admitting analytical solutions for which the continuity
conditions can be easily solved with high accuracy, whereas solving
numerically the Schr\"odinger equation is very uncertain at low energies.
The calculations we present are performed for $a=1$. The features we emphasize in
this work are not qualitatively affected by this parameter.
%end
From the numerical results, we get the following approximate
asymptotic laws :
\be\label{13}
\langle r^2 \rangle_0 \approx {c_0 \over {|E_0|}}~~~~
\langle r^2 \rangle_1 \approx {c_1 R_{cut} \over 
 { \sqrt{ \vert {E_1} \vert}}}~~~~
\langle r^2 \rangle_2 \approx {c_2 {R_{cut}^2}} \ . 
\ee
Here $c_1$ and $c_2$ slightly depend on $R_{cut}$, whereas $c_0 = 1/2$ as
expected (\ref{5}).

Relations (\ref{13}) show that, for any $\ell$,
the dependence
of the mean square radii on the energy of the state 
is not modified even in presence of a core. This is in agreement with what
the arguments based on the asymptotic
behaviour of the wave functions would suggest (\ref{8}).

Concerning the dependence of the mean square radii on $R_{cut}$,
we see from (\ref{13}) that the cut Kratzer potential present  a {\it quasi}-scaling,
the dependence
on the range of the potential $R_{cut}$ being
very similar to that of
(\ref{8}). 

In order to show the dependence of the $c_{\ell}$ on $R_{cut}$
and the shape of the potential, 
these coefficients 
are shown, as a fonction of the energy,
in figure 1 (up $\ell=0$, middle  $\ell=1$, bottom  $\ell=2$),
 for $R_{cut}=3$ (line), $6$ (short-dashed line) and 
$12$ (long-dashed line) . For comparison we show 
the  $c_{\ell}$ obtained for a short range potential defined by two
parameters (dotted line). We take the
square well as an example. 

We see that, 
apart from the $\ell = 0$ case, as the energy tends to zero, the asymptotic
values of $c_{\ell}$ depend indeed on the potential. This dependence
increases with $\ell$, as one should expect if the centrifugal
barrier has a confining effect and the wavefunctions become more
sensitive
to the particular shape of the potential. 
We also see that for $\ell=1,2 $ 
the spreading among the three cut Kratzer
potential results is only of a few percents (legitimating the {\it quasi}-scaling), 
much less than the differences with
the square well potential results. 
This qualitatively underlines the
differences in the inner part of the wavefunctions. 
%Note that the smallness
%of the spreading of the results obtained with the cut Kratzer potentials
%legitimates the concept of quasi-scaling. However, one has to keep in mind
%that this conclusion relies on specific numerical examples, and may not be
%general. 

%%%%%%%%%%%%%%%%%
It is interesting to ask if specific physical systems are in the
low-energy limit where the asymptotic laws directly give the size of
the system once the separation energy is known.
We have looked at the case of halo nuclei and molecular dimers, by using
experimental values of the separation energy and radius to calculate the
corresponding $c_0$.
For $\rm  ^{11}Be$ the experimental separation energy is $.5~ MeV$ and the rms
radius is about $7~fm$, which leads to $c_0 \approx 1.1$. The
$\rm ^4He - ^4He$ dimer has a separation energy of $10^{-1} \mu eV$ and a
radius of 70 {\AA} \cite{STO}
; it yields $c_0 = .52$. The calculated $\rm ^3He - ^6Li$
\cite{KLM}, with a separation energy of $10^{-2} \mu eV$ and a radius of 210
{\AA}, reaches $c_0 = .5$.  These estimates show that
the loosely bound  dimers are
situated in the asymptotic 
region, whereas the halo nuclei merely remain at its
threshold, in a range sensitive to the potential. A similar conclusion
has been drawn by studying the Bertlmann-Martin inequality \cite{RJL}.

Outside the asymptotic region, it becomes even difficult to decide between a
short and a long range potential, if the only observables are the separation
energy and the rms radius. The key point in this respect is obviously given
by the spectrum. In both dimers and halo nuclei, the number of states is
rather limited, and thus a description by a short range potential is quite
natural.

Let us now  come to the
transition between the short and long
range regime. The example of the 
Kratzer potential provides us with a way of studying 
this transition  by looking at evolution of $c_{\ell}$ 
as a function of the range
$R_{cut}$. 
A priori, the transition depends on the energy, but  
numerical results show that this dependence is
simple in our case. In figure 2 we present
$c_0$ against
$\sqrt{|E_0|} R_{cut}$. 
The half-point between the short range ($c_0=1/2$) and long range 
regime ($c_0=3$)
lies at around $\sqrt{|E_0|} R_{cut} = 2$.

For higher angular momenta, the problem is more complicated 
because going from the short (\ref{12}) to the long range (\ref{13})
regime 
modifies both
$c_{\ell}$ and the energy dependence of the asymptotic laws.

Note that the scaling and {\it quasi}-scaling present in
eqs.(\ref{8}) and (\ref{13}) cannot be used to study the transition
between short and long range regime.
For example, for $\ell =0$,
the transition does not occur if we take the square well potential at any
finite radius $R_0$. This is due to the fact that according to eq.(\ref{8}-\ref{5})
$c_0= 1/2$ independently on $R_0$. Nevertheless, it is clear that
for a fixed enegy, the required potential deepness $\lambda$ is decreasing
as $R_0$ increases, so that the limit $R_0 \rightarrow \infty$ implies
$\lambda \rightarrow 0$. It means that the square well potential only
reaches the long range limit for infinite $R_0$.

%On one hand, it
%is possible to find, for a given energy, the cutting radius necessary to
%obtain the long range 
%value of the $<r^2>_{\ell}$. However, $c_{\ell}$ varies with
%$R_{cut}$, and the few values we have calculated 
%are not sufficient to draw general rules, specially if the required 
%$R_{cut}$ becomes large.

Let us finally remark that up to now the weakly bound objects 
found 
experimentally are intepreted in the two-body description
as s-states. It remains a
challenging problem to observe  halo states that would
be interpreted, in this context, 
as higher angular momentum states.
To the extent that a two-body description yields the basic description,
they should occur as excited levels. In the nuclear case, this situation is
hardly possible, because the many-body degrees of freedom will strongly
affect the first $s$-state, and destroy the simple 
two-body picture.  From this point of view, the dimers are better candidates.

In conclusion, we have discussed the size of two-body weakly bound
objects using both long range potentials and short range potentials
having a repulsive core. We have shown that not only short range
but also long range potentials give rise, in the zero energy limit,
to very extended systems. The asymptotic
laws obtained in this case
show a different energy dependence from those of short range
potentials already for p- and d-states. 
The dependence of the asymptotic laws
on the particular shape and typical length of short range
potentials, defined by both two and three parameters, has also
been studied. 
All these ideas
as well as the transition of the asymptotic 
laws 
from the short to the long range regime for $\ell=0$ are illustrated in the
particular case of the Kratzer potential.

\vspace{0.2cm}
\noindent
{\bf Acknowledgements}

\noindent
We express our thanks to M. Lassaut for helpful discussions.

\newpage

\begin{figure}
\begin{center}
\includegraphics[angle=-90.,scale=0.6]{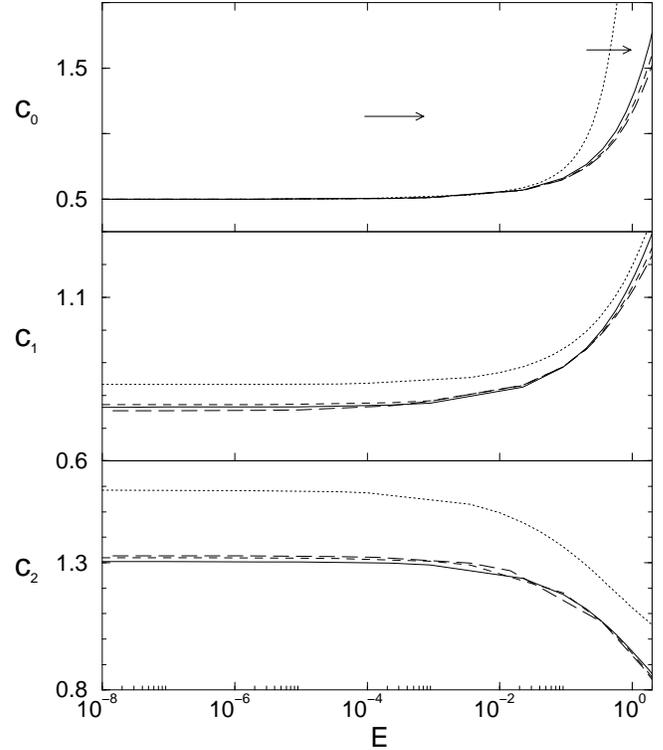}
\end{center}
\protect\caption{cut Kratzer potential : 
the values of
$c_{\ell}$ are plotted against $E_{\ell}$ for $\ell$ = 0 (up), 1
(middle) and 2 (bottom).
The 
line corresponds to $R_{cut}$ = 3; the short-dashed and long-dashed  
lines correspond to
$R_{cut}$ = 6 and 12, respectively. 
For comparison, the dotted line shows the results obtained with a
short
range potential without the repulsive core. As an example,
the square well is taken.
As far as $c_0$ is concerned, the arrows indicate  the range of the lowest
energies corresponding to halo nuclei (arrow on the right) and dimers 
(arrow on the left).}
\end{figure}

\begin{figure}
\begin{center}
\includegraphics[angle=-90.,scale=0.6]{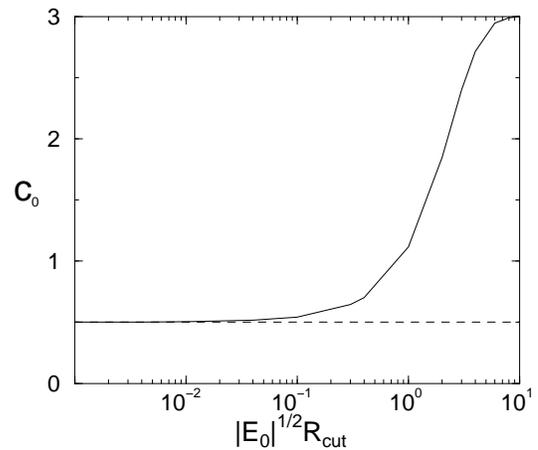}
\end{center}
\protect\caption{Evolution of $c_0$ from the short to the long range
  regime for the cut Kratzer potential. 
For comparison,
the dashed line shows the value of 1/2 valid when $E_0 \rightarrow 0$
for any short range potential (\ref{5}).}
%%%%%The value of $c_0$ is plotted against $\sqrt{|E_0|} R_{cut}$.}
\end{figure}

\end{document}